\def\beg{\begin{equation}}
\def\eeq{\end{equation}}
\documentstyle[12pt]{article}
\begin{document}
\begin{center}
{\Large{\bf Violation of classical electrodynamics by composite fermions in the quantum Hall effect.}}
\vskip0.35cm
{\bf Keshav N. Shrivastava}
\vskip0.25cm
{\it School of Physics, University of Hyderabad,\\
Hyderabad  500046, India}
\end{center}

We find that the composite fermion (CF) which is the magnetic flux quanta attached to the electron, although based on experimentally observed fractions in the quantum Hall effect, is inconsistent with
the classical electrodynamics. It produces new quasiparticles which
have the usual charge and spin but their rotation in a curved path
does not produce magnetic field. This ``zero-field" production is inconsistent with the classical electrodynamics. The attaching and detaching of magnetic flux quanta to the electron, also violates the theory of relativity.
\vfill
Corresponding author: keshav@mailaps.org\\
Fax: +91-40-3010145.Phone: 3010811.
\newpage
\baselineskip22pt
\noindent {\bf 1.~ Introduction}

     When a metal sheet is placed in a magnetic field along the $z$ direction with the applied voltage along the $y$ direction, there
is a current in the $x$ direction. This current is called the Hall current. The $x$ component of the resistivity is the Hall resistivity which varies linearly with the applied field along z direction.
In 1980 it was found that there is a plateau in the Hall 
resistivity in the $xy$ plane at the value of $\rho=h/ie^2$ where 
i is an integer. Later on it was found that a plateau in the transverse resistivity also occurs at $i$=1/3 so that this number need not be an integer.(We now believe that this quantization of resistivity is the result of flux quantization, and not due to ``fine structure"
as was thought in the original work). 
     In 1989, Jain$^{1,2}$ constructed the denominator, 3 by using 2$\pm$1 so that one of the fractions is 1/(2+1) = 1/3 and an even
 number, 2 is needed. It is proposed that there are two types of quasiprticles, the electrons and the composite fermions (CF). The magnetic field at the CF is less than at the electron,
\beg
B^* = B - 2mn\phi_o
\eeq
where m is an integer, n is the number of electrons per unit area and $\phi_o$ is the unit flux quantum, hc/e, with h as the Planck's constant, c the velocity of light and e, the charge of the electron.
 The number 2 may be written as 2m and the filling factor, 
\beg
\nu = {p\over 2mp + 1}.
\eeq
For m=1, $\nu$=p/(2p+1). When compared with the experimental data,
 the series of fractional charges, p/(2p+1) is found to be correct
 and there is no doubt that this is the correct series. In the
 begining of this subject, it was thought that the denominators
 are odd numbers in the series which gives the fractional charge, i.e.,
\beg
\nu = {1\over 2p+1}.
\eeq
Laughlin obtained the wave function of a quasiparticle of charge
$e_{eff}$=(1/3)e. There is nothing against the quantum mechanics
 which teaches us to write wave functions but in quantum Hall effect
it is the measurement of current and voltage which plays the dominant role. An examination of the expression (1) shows that even number
of flux quanta are attached to the electron. Let us examine, if this
attachment of flux quanta is consistent with the classical electrodynamics.

\noindent{\bf 2.~~Explanations}

    The Biot and Savart's law shows that when an electron travels
in a circular orbit of radius, $R$, it produces a magnetic field.
 If the orbit is in the $xy$ plane, then the field produced is 
in the $z$ direction. One electron produces a field of B and the 
flux is quantized so that,
\beg
B={n\phi_o\over A}
\eeq
where $A$ is the area, which we can calculate if we know B. If 
2$n\phi_o$ is attached to one electron, the field will be given 
by B$^*$ instead of B where
\beg
B^* = B - 2n\phi_o.
\eeq
When the field B comes from the charge $e$, what happened to the
 charge corresponding to 2$n\phi_o$? Therefore, we understand that 2$n\phi_o$ is decoupled from the charge. The current I corresponds to field B but 2$n\phi_o$,
\beg
B={2I\over cR}
\eeq
where c is the velocity of light. We show the current I in Fig.1
 which gives the field B so that I gives B but not B$^*$. To get 2$n\phi_o$, we detach it from I$_2$ and give it to B. Thus
 we have the CF which has a current of I and the field of 
B+2$n\phi_o$ and the current I$_2$ with zero field. Thus we have
 two new quasiparticles, one of these is called the composite
 fermion (CF) and the other, a new quasiparticle, called the 
decomposite fermion (DF). What is the CF? {\it The CF is the 
electron current I with field B to which an extra field has 
been added}. Now, what is called DF? {\it The DF is the current
 I$_2$ from which the field has been detached}. Thus the composite fermion has a field of B+2$n\phi_o$ and the decomposite fermion 
has a field of zero. In the classical electrodynamics, the current 
I gives the field B but the CF has a field of B+2$n\phi_o$ and
 the DF has zero field. We have thus invented two new 
quasiparticles, the CF and the DF which are not the same as 
electrons. The CF was suggested by Jain in 1989 to explain the 
quantum Hall effect and the DF has been found only in the present
 work. It is clear that both the CF as well as the DF are not 
consistent with the classical electrodynamics. The construction of
 a magnet is based on the Biot and Savart's law which is believed
 to be correct. Therefore both the CF and the DF are``incorrect". Dyakonov$^3$ has said that although, many experimentalists support Jain's idea of CF moving in a field, NO BODY HAS SHOWN
 THEORETICALLY THE EXISTENCE OF CF AS (QUASI)FREE PARTICLES. 
Farid$^4$ also pointed out that the reduced field formula 
is not correct.\\

\noindent{\bf3. ~~ The decomposite fermion (DF)}

     The decomposite fermion (DF) is created when CF is formed. 
Since the magnetic field of 2$n\phi_o$ has been detached from the electon, what is left is a particle with spin and orbit but no 
magnetic field. Once the magnetic field has been detached from the electron it will not emit light. Similarly, light signals will not 
reach it. This will be a dark object and there will be no way of 
seeing it. No electromagnetic signal will reach it nor will it 
emit any. However, it has a finite mass so that formation of CF will result into overall reduction of mass. Since some of the electrons 
will be lost by detachment of magnetic fields, there is overall loss
 of mass. The DF will stay in a dark region. The GaAs is 
electrically neutral. When charge goes to the dark region, the
 GaAs will become positively charged. In this way large positive 
charge will collect on the GaAs and a current will flow due to 
formation of CF.

     As a further clarification, the electrons in a curved path
 produce a field. When this field is removed, what is left is a 
particle of charge $e$ and spin $s$ but no magnetic field. This 
particle is called the detached fermion (DF). This quasiparticle
 is not consistent with the classical electrodynamics. Since the magnetic vector has been detached from the spin and chartge, the
 DF will  not be consistent with the hypotheses of the ``special
 theory of relativity."

\noindent{\bf4.~~ Comments on CF}.

     We have shown$^5$ that the CF is a large object and hence
 it is impossible that it has the same density as the electron. 
The CF model of Jain$^{1,2}$ which is used to understand the 
quantum Hall effect requires that the density of CF should be 
the same as that of the electron. Therefore, the CF model is
 internally inconsistent. We have shown$^6$ that the expression 
for the effective field at the CF site is not correct. We have 
shown$^7$ that the phenomenology of the CF is not correct. The experimentalists have been misguided to claim that CF has been 
observed. In fact$^4$, the observation  has nothing to do with
 the CF. The CF should be a fermion by definition but it is found
 that they cease to be fermions$^9$.

In fact all of the experimental data published in the PRL 1998-2001 is in agreement with the angular momentum model$^{10}$.
\vskip0.25cm
\noindent{\bf 5.~~Conclusions}

     The CF model$^{1,2}$ of the quantum Hall effect is internally inconsistent. It violates the laws of the classical electrodynamics.
 The formation of the CFs requires that there should exist DFs. Thus flux quanta are attached to the electron or detached from the 
electron. In the later case, it is clear that the CF violates the "special theory of relativity." Kukushkin et al$^{11}$ were misguided
 to make a claim that they observed "flux attached" CFs. These authors$^{11}$ actually did not observe the CFs. What they observed is not a CF.

     The correct theory of the quantum Hall effect is given in ref.12.

\noindent{\bf6.~~References}
\begin{enumerate}
\item J.K. Jain, Phys. Rev. Lett. {\bf63}, 199 (1989).
\item K. Park and J.K. Jain, Phys. Rev. Lett. {\bf81}, 4200 (1998).
\item M. I. Dyakonov, cond-mat/0209206.
\item B. Farid, cond-mat/0003064.
\item K. N. Shrivastava, cond-mat/0209057.
\item K. N. Shrivastava, cond-mat/0207391.
\item K. N. Shrivastava, cond-mat/0204627.
\item K. N. Shrivastava, cond-mat/0202459.
\item K. N. Shrivastava, cond-mat/0105559.
\item K. N. Shrivastava, cond-mat/0201232.
\item I. V. Kukushkin et al, Nature {\bf415}, 409 (2002).
\item K.N. Shrivastava, Introduction to quantum Hall effect,\\ 
      Nova Science Pub. Inc., N. Y. (2002).
\end{enumerate}
\vskip0.1cm
Note: Ref.12 is available from:\\
 Nova Science Publishers, Inc.,\\
400 Oser Avenue, Suite 1600,\\
 Hauppauge, N. Y.. 11788-3619,\\
Tel.(631)-231-7269, Fax: (631)-231-8175,\\
 ISBN 1-59033-419-1 US$\$69$.\\
E-mail: novascience@Earthlink.net

\vskip0.5cm

Fig.1: The current I produces a field B where the radius of curvature is R. This is called the Biot and Savart's law. The current I$_2$ produces a field 2$n\phi_o$. This 2$n\phi_o$ is detached from I$_2$ and attached to I to make a field of B+2$n\phi_o$.
Thus the quasiparticle which has a field of B+2$n\phi_o$ is called CF. This leaves a particle of current I$_2$ with zero field called DF. We show that CF-DF model is inconsistent with classical electrodynamics
 and relativity.
\end{document}